\documentclass[aps,pre,preprint,double-spaced,groupedaddress]{revtex4-1}
\usepackage{amsmath,amsthm,amssymb,graphicx}
\title{\LaTeX\\Sufficient Condition for Presence of Spontaneous Magnetization on a General Graph}
\date{13/11/2009}
\newcommand{\G}[3]{\ensuremath{ {\cal G{#3}}^{#2}_{#1} } }
\newcommand{\Gg}{\ensuremath{ {\cal G} } }
\newcommand{\Gn}[1]{\G{N}{}{#1}}

\newcommand{\Z}{\ensuremath{{\cal Z}}}

\newcommand{\boltz}{ \ensuremath{e^{-\beta{\cal H}}} }
\newtheorem*{definition}{Definition}

\newtheorem*{theorem}{Theorem}

\newtheorem*{proposition}{Proposition}

\begin{document}
\title{Generalization of the Peierls-Griffiths Theorem for the Ising Model on Graphs}
\author{Riccardo Campari}
\email{riccardo.campari@fis.unipr.it}
\author{Davide Cassi}
\email{davide.cassi@fis.unipr.it}
\affiliation{Dipartimento di Fisica, Universit\`a di Parma, Viale Usberti 7/A, 43100 Parma, Italy}

\date{\today}

\begin{abstract}
We present a sufficient condition for the presence of spontaneous magnetization for the Ising model on a general graph, related to its long-range topology.
Applying this condition we are able to prove the existence of a phase transition at temperature $T>0$ on a wide class of general networks.
The possibility of further extensions of our results is discussed.
\end{abstract}

\maketitle

\section*{Introduction}
Since the original paper by Ising\cite{Ising1925}, in which it was proved that the Ising model on an infinite linear chain would not show a phase transition, a huge amount of research has been conducted on the subject. The first phase of this fruitful line investigated regular systems, and after the articles by Peierls\cite{Peierls1936} and Onsager\cite{Onsager1944} it became clear that regular lattices in $d$ dimensions would magnetize when $d\ge2$.\par
In a second phase, a large number of fractals was investigated\cite{YevalGefen1980,MonHsi2003,Achiam1985,CarMarRui1998,BaFaAl2008}, mainly via the so-called renormalization group techniques, to discover that, although no rigorous theorem has been proven,
those (and only those) fractals which have an infinite minimum order of ramification display spontaneous magnetization.\par
In the same years, fundamental analytical results were obtained for disordered structures embedded in Euclidean lattices, applying percolation theory concepts\cite{ChaFro1985}.\par
More general graphs\cite{Jullien1979,ReggeZecchina1996,Ceresole1998,Herrero2002,DogGolMen2002} have become increasingly popular in the last twenty years:
the main difference with the previous cases is that the metric structure
of the embedding space ceases to play an essential role, as in general a graph is a topological structure which is not necessarily embeddable in a finite dimensional Euclidean space.
The absence of translational invariance and scale invariance makes general graphs very difficult to study, as \emph{ad hoc} techniques must be employed, that usually admit no straightforward generalization.\par
An important result would be the identification of a simple parameter, capable of determining whether the Ising model on a given graph exhibits a phase transition: we present here a theorem stating a sufficient condition for a graph to exhibit spontaneous magnetization, which is a generalization of the classic Peierls-Griffiths theorem\cite{Peierls1936,griffithspeierls} for the square lattice.
While interesting works, employing the same basic techniques as the Griffiths theorem, have been proposed for higher-dimensional lattices, typically stemming from the paper of Dobrushin\cite{Dobrushin1964rus,Dobrushin1964}, such as the profound contribution by Isakov\cite{Isakov1984} and the extensions to non-symmetric situations treated in Pirogov-Sinai theory\cite{Sinai1982}, or directly from the paper by Griffiths, as in Lebowitz and Mazel\cite{Lebowitz1998}, nothing applying to inhomogeneous networks and arbitrary graphs has yet emerged, and our contribute aims essentially at filling this gap.\par
The reason why the modulus of magnetization $\langle \left\vert M\right\vert\rangle$ is considered is that it is indissolubly tied to the long range order of the graph: it is easy to prove that, when the external field is zero, stating $\langle \left\vert M\right\vert\rangle \;\ge \epsilon >0$ is equivalent to the existence of a non-zero measure subset of all the correlation functions such that all of its members are greater than a small constant $\epsilon'>0$.\par
In the following, we first present the concepts of open and closed borders in a graph for later use; we then define the ferromagnetic Ising model on a general graph and derive the equivalence of the sum over configurations and the sum over different borders. Next we prove a theorem stating a sufficient condition for a graph to exhibit spontaneous magnetization. Because of the technical nature of the theorem, we thoroughly examine its more and less immediate consequences for a wide range of different graphs. Lastly, we discuss our results and the current comprehension of the mechanism of spontaneous magnetization on graphs for the Ising model.

\section*{Open and Closed Borders in a Graph}
A {\bf graph} $\cal G$ is a pair $\left({\cal P},{\cal L}\right)$, where $\cal P$ is a countable collection of vertices and ${\cal L}\subset {\cal P}\times{\cal P}$ is a set of unoriented bonds between points.
Any pair ${\cal G}'=({\cal P}',{\cal L}')$, such that ${\cal P}'\subset {\cal P}$, ${\cal L}'$ contains only links between elements of ${\cal P}'$ and ${\cal L}'\subset {\cal L}$, is called a {\bf subgraph}, and it's denoted ${\cal G}'\subset {\cal G}$.
We will restrict our attention to those graphs whose coordination number $z_i$, representing the number of bonds in ${\cal L}$ having one extremum in $i$, is uniformly limited: an integer $z_{Max}>0$ exists such that $z_i\le z_{Max}$ for all $i\in{\cal P}$.\par
We now define a {\bf path} $\gamma$, between two points $i$ and $j$, as a collection of consecutive bonds of $\cal L$, 
where consecutive means that each pair shares a vertex with the next one:
\[ \gamma = \left\{ (il_1), (l_1 l_2), \dots , (l_{D-1},j) \right\}. \]
Directly associated to the concept of path, the \textbf{chemical distance} between two points $i$ and $j$ is defined as the length of the shortest path connecting them. The \emph{chemical distance} straightforwardly induces the so-called intrinsic metric of the graph.

The \textbf{intrinsic fractal dimension} $d_{frac}$ of a graph, defined as the minimum $d$ such that $N_r$, the maximum number of vertices included in a Van Hove sphere\cite{BurioniRWAG} of radius $r$ (i.e. the set of points within a chemical distance from a given point of no more than $r$ bonds), satisfies $N_r\le r^{d}$ as $r\rightarrow\infty$. It differs from the usual \emph{fractal dimension} in that it refers to the topological nature of the graph (i.e. on its natural - chemical - distance), and not on the metric structure of the space into which the graph is embedded.

To proceed we declare what will be considered a border from now on.
\begin{definition}
Given a connected graph $\Gg = \left( {\cal P}, {\cal L} \right) $,
we can define a {\bf border} $\cal B$ as a set of bonds that separates exactly two connected subgraphs. It means that two sets ${\cal P}_1,{\cal P}_2\subset{\cal P}$ exist such that
\begin{itemize}
\item ${\cal P}_1 \cap {\cal P}_2 =\emptyset$ and ${\cal P}_1 \cup {\cal P}_2 = {\cal P}$,
\item any path on \Gg from a point of ${\cal P}_1$ to a point of ${\cal P}_2$ must contain at least one bond of $\cal B$,
\item a path exists between any two points in ${\cal P}_i$ $(i = 1,2)$ that doesn't contain any bond of $\cal B$.
\end{itemize}
\end{definition}

It is noteworthy that the union of two disjoint borders is not a border itself under this definition, as it divides the graph into three subgraphs. This is a feature we'll later need to avoid overcounting different configurations.\par
The intuitive idea of open and closed border
is actually an artifact created by our visualizing regular lattices as immersed in a finite dimensional real space: the seeming adjacency of the vertices creates a contour of the graph, which we use to define closed and open borders. The fact is that this contour is heavily dependent on what particular immersion we employ, and ceases to exist when we consider the graph for itself.
The border in itself has no geometry whatsoever, since it is just a collection of links, and even the notion of "continuous" border, without further specifications, makes no sense from a graph-theoretic point of view: in a general graph, a border is just a collection of links that splits it into two parts.
We now define open and closed borders with respect to an external set of points, as it will be useful later.
\begin{definition}
Given a border $\cal B$ and a set of points ${\cal E}\subset{\cal P}$, we say that $\cal B$ is {\bf closed} with respect to the \emph{external points set} $\cal E$ if either ${\cal P}_1\cap{\cal E}=\emptyset$  or ${\cal P}_2\cap{\cal E}=\emptyset$, otherwise $\cal B$ is {\bf open}.
\end{definition}

For any finite subgraph ${\cal G}_N$ of a given graph $\cal G$, we choose the natural set of external points $\cal E$:
\begin{eqnarray*}
{\cal E} \equiv \left\{ i\in\Gn{} :  (i,j)\in{\cal L}\;\; for\;\, some\,\; j\in\Gg\setminus\Gn{}  \right\}.
\end{eqnarray*}
Now, given a border ${\cal B}_i$ that divides \Gn{} into two subgraphs $A_i$ and $C_i$, we define $A_i$ as
\begin{itemize}
\item \emph{internal} if
\begin{itemize}
\item ${\cal B}_i$ is closed and $A_i\cap{\cal E}=\emptyset$, or
\item ${\cal B}_i$ is open and $A_i$ contains fewer elements than $C_i$, or
\item ${\cal B}_i$ is open, $A_i$ has the same size as $C_i$ and the points in $A_i$ linked to ${\cal B}_i$ have negative spin;
\end{itemize}
\item \emph{external} if
\begin{itemize}
\item ${\cal B}_i$ is closed and $A_i\cap{\cal E}\ne\emptyset$, or
\item ${\cal B}_i$ is open and $A_i$ has more elements than $C_i$, or
\item ${\cal B}_i$ is open, $A_i$ has the same size as $C_i$ and the points in $A_i$ linked to ${\cal B}_i$ have positive spin.
\end{itemize}
\end{itemize}
The reason why we had to select a finite subgraph ${\cal G}_N$ is that we need to be able to count the number of spins in the graph for the previous definitions to make sense.

\section*{The Ferromagnetic Ising Model on a Graph}
Let now $\sigma_i=\pm 1$ be a spin variable for each vertex $i \in {\cal P}$.
We define the {\bf Ising Hamiltonian} on a graph as
\begin{equation}
{\cal H} = -\sum_{(i,j)\in {\cal P}\times{\cal P}} J_{ij} \sigma_i \sigma_j - \sum_{i\in {\cal P}} \sigma_i h_i,
\end{equation}
where the couplings $J_{ij}=J_{ji}$ must satisfy $0\le J_{ij} < J_{Max}<\infty$ for some $J_{Max}$, and $J_{ij}>0$ if and only if $(i,j)\in{\cal L}$.
In the following we will set the external field to zero everywhere ($h_i\equiv 0$).\par
Now that we have presented the terminology we'll be using, we are going to study the equilibrium statistical mechanics of the Ising model at inverse temperature $\beta$, and in particular the modulus of the magnetization
\begin{equation}
\langle |M|\rangle = {\cal Z}^{-1} \sum_{\left\{ \sigma_i \right\} } { \left\vert \sum_{j\in{\cal P}} \sigma_j \right\vert \over \left\vert{\cal P}\right\vert } e^{-\beta {\cal H}(\left\{ \sigma_i \right\})},
\end{equation}
where ${\cal Z} = \sum_{\left\{ \sigma_i \right\} } e^{-\beta {\cal H}(\left\{ \sigma_i \right\})}$ is the partition function, and $\left\vert {\cal P} \right\vert$ is the cardinality of $\cal P$.\par
Since $\Delta |M| = \langle |M|^2\rangle - \langle |M|\rangle\langle |M|\rangle\;$ is a variance,
\[\langle M^2\rangle \;\ge (\langle |M|\rangle)^2,\]
so stating $\langle |M|\rangle\;=\epsilon>0$ implies $\langle M^2\rangle\;\ge \epsilon^2>0$. On the other hand, since $M^2\le |M|$, the converse is true, so $\langle M^2\rangle\;>0 $ and $\langle |M|\rangle\;>0$ are equivalent.

As we have now defined the main quantities we'll be studying, our next step is to prove that we can substitute the sum over configurations of the graph with a sum over possible \emph{border classes}, that we now define.

\section*{Equivalence between sets of borders and spin configurations}
\begin{definition}
A {\bf border class} is a class $C=\{C^i\}$ of border sets $C^i = \left\{ B^i_1,B^i_2,\dots,B^i_{N_i} \right\}$, where $i=1,\dots,N_C$, such that
\begin{itemize}
\item $B^i_u\cap B^i_v = \emptyset$ for all $i=1,\dots,N_C$ and $u,v=1,\dots,N_i$,
\item $\cup^{l=1,\dots,N_i} B^i_l = \cup^{m=1,\dots,N_j} B^j_m$ for all $i,j=1,\dots,N_C$.
\end{itemize}
\end{definition}
\begin{theorem}
To any given border class corresponds one and only one configuration of spins on $\cal P$, once we set the value of a single spin.
\end{theorem}
\begin{proof}
To prove that, for any border class and a given spin $p\in{\cal P}$, we can construct a single spin configuration, we first choose an arbitrary representative $C^i = \left\{ B^i_1,\dots,B^i_N \right\}$ of $C$ and set all the spins to the value of $p$, then for each $B^i_k \in C^i$ we flip all the spins of the subgraph which doesn't contain $p$. The result is independent of the order in which we choose the $B^i_k$, since each spins changes sign once for every border that separates it from the fixed spin $p$, and is independent of the specific $i$.\\
To prove that for any given spin configuration we can create a single border class, we proceed as follows: let $R^{\pm}$ be the sets of all plus (minus) spins,
\begin{eqnarray*}
R^{\pm} \equiv \left\{ i\in\Gg : \sigma_i=\pm 1 \right\};
\end{eqnarray*}
We now choose the subsets $R^\pm_i$ of $R^{\pm}$, so that each $R^\pm_i$ is connected, while for all $i\ne j$ $R^\pm_i$ and $R^\pm_j$ are disconnected; moreover we require that
\begin{eqnarray*}
&&R^\pm \equiv R^\pm_1 \cup R^\pm_2 \cup \dots \cup R^\pm_S,\\
&&R^\pm_i\cap R^\pm_j = \emptyset \;\;\;\forall_{i\ne j}.
\end{eqnarray*}
We are selecting individual clusters of homogeneous spins, so satisfying the above requisites is always possible.
Setting now
\begin{eqnarray*}
\partial R^\pm_i \equiv \left\{ (a,b)\in{\cal L} : a\in R^\pm_i, b\notin R^\pm_i \right\},
\end{eqnarray*}
the sets $B^\pm \equiv \partial R^\pm_1 \cup \dots \cup \partial R^\pm_S$ are a collection of links each defined unambiguously, and furthermore $B^+\equiv B^-$.
It may happen that for some $i$ the subgraph $\Gg \setminus R^\pm_i$ is made of two disconnected subgraphs (e.g. when a ring of plus spins is surrounded by minus spins); as a consequence $\partial R^\pm_i$ is not a border according to our definition. In that case it is possible to split $\partial R^\pm_i$ into subsets, so that each of them divides \Gg into two connected subgraphs. After dealing in this way whith all the $R^\pm_i$, we are left with a collection of well-defined borders $\partial T^+_j$, with $j=1,\dots,U_+$, and $\partial T^-_k$, with $k=1,\dots,U_-$ where $U_+,U_->S$.
It is still possible that some of the borders $T^+_i$, while defining exactly the same zones, have no correspective in $T^-_i$ but, since they nevertheless verify $B^+\equiv B^-$, they belong to the same border class, completing the proof.
\end{proof}
The main consequence of this result is that we can substitute a sum over border classes for a sum over configurations whenever needed, and we can infer from the structure of the borders some limiting properties for the spins distributions, as we'll see soon.
It is worthwhile to explicitly notice that, when we pass from a sum over configurations to one over borders, and not border classes, we overcount some borders, as there are more than one representative of each border class: this is not going to be a problem in the use we'll make of this result.

\section*{Generalized Peierls-Griffiths' Theorem}
We can divide the set of all configurations on \Gn{} into two classes:
\begin{itemize}
\item all the negative spins are internal to some border (class $\cal N$),
\item at least a negative spin exists that is external to all borders (class $\cal P$).
\end{itemize}
The second case implies that every positive spin lies inside some border, since it must lie on the opposite side of the negative spin which is always external.\par
We now restrict our attention to the configurations belonging to the first class, denoting by a subscript $\cal N$ the quantities that pertain to it; we can obtain a good estimate of the number of negative spins, $\langle N_-\rangle_{\cal N}$, as follows: the sign of a spin $p$ is negative if it is contained inside an odd number of borders, positive otherwise; we obtain a very naive, yet effective, approximation if we consider any spin contained inside at least one border as negative: letting $I^p$ be $1$ if $p$ is inside at least a border, $0$ otherwise, we can write
\begin{eqnarray*}
\langle N_-\rangle_{\cal N}\;\; \le \sum_{p\in{\cal G}} \;\langle I^p\rangle.
\end{eqnarray*}
We are now to give a reasonable estimate of $\langle I^p\rangle$: take all the configurations $\cal C$ with at least one border containing $p$, call $b_{min}$ the length of the shortest border in $\cal C$ containing $p$ and let $k$ be the number of borders containing $p$, so as to write
\begin{eqnarray*}
\langle I^p\rangle\;\; = \Z^{-1}\sum_{ {\cal C}|_{p\;inside} } \boltz = \Z^{-1}\sum_{b_{min} \ge 1} \sum_{k \ge 1} \sum_{{\cal C}|^{b_{min}}_{k\;borders}} \boltz.
\end{eqnarray*}
Now fix $b_{min}$ and consider the configurations containing $k$ borders: if we remove the shortest border from such a configuration $\cal C$, we obtain a new configuration ${\cal C}'$ with $(k-1)$ borders containing $p$, each of them at least $b_{min}$ long. ${\cal C}'$ will be present in the partition function $\cal Z$, but different configurations with $k$ borders $\cal C$ may give the same ${\cal C}'$;
defining now $\mu^p(b)$ as the number of possible borders of $b$ links containing $p$, the degeneration induced by removing the shortest border is not greater than $\mu^p(b_{min})$.
The energy of a configuration $\cal C$ and the corresponding Boltzmann factor obey
\begin{eqnarray*}
&&E_{\cal C} \ge E_{{\cal C}'} + 2\beta J_{min} b_{min},\\
&&e^{-\beta{\cal H}_{{\cal C}}} \le e^{-{\cal H}_{{\cal C}'}}e^{-2\beta J_{min} b_{min}};
\end{eqnarray*}
if we now limit the sum in the partition functions to those configurations obtained by removing a border from the numerator, we can write
\begin{eqnarray*}
{\cal Z}^{-1} \sum_{k \ge 1} \sum_{{\cal C}|_{k\; borders}} \boltz \le \mu^p(b_{min})\; e^{-2\beta J_{min} b_{min}}.
\end{eqnarray*}
In this way the average number of minus spins is bounded by a function depending only on the number of borders encircling a given spin:
\begin{eqnarray*}
\langle N_-\rangle_{\cal N}\;\; \le \sum_{p\in{\cal G}} \sum_{b_{min}} \mu^p(b_{min})\;e^{-2 \beta J_{min} b_{min}};
\end{eqnarray*}
this result states that, no matter what the maximum number of spins you can isolate inside a border is, as long as $\mu^p(b)$ grows at most exponentially the value of $\langle N_-\rangle_{\cal N}$ can be limited at low enough temperatures.
An analogous result holds for configurations of class $\cal P$ when exchanging the roles of positive and negative spins:
\[ \langle N_+\rangle_{\cal P}\;\; \le \sum_{p\in{\cal G}} \sum_{b_{min}} \mu^p(b_{min})\;e^{-2 \beta J_{min} b_{min}}. \]

Let now $\mu(b)) = \sup_p\mu^p(b)$. We can now prove the following theorem:
\begin{theorem}
If on an infinite graph $\mu(b)\le A^{b}$, definitely for $b\ge\bar{b}$ and for some for some $A>0$, then the graph exhibits spontaneous magnetization at large enough $\beta$ (low enough temperatures).
\end{theorem}
\begin{proof}
The average modulus of magnetization is $|M| = N^{-1}\left( N_+-N_- \right)$; writing the Boltzmann factor for a configuration $\cal C$ as $P_{\cal C} = {\cal Z}^{-1} e^{-\beta {\cal H}({ \cal C })}$, we can write the following:
\begin{multline*}
\langle |M|\rangle\;\;=\;\sum_{\cal C \in P \cup N}\;|M_{\cal C}|\;P_{\cal C} \;= \sum_{\cal C \in (P \cup N)^+} M_{\cal C}\;P_{\cal C} - \sum_{\cal C \in (P \cup N)^-} M_{\cal C}\;P_{\cal C}\\
= \left( \sum_{\cal C \in N^+}\;M_{\cal C} P_{\cal C} - \sum_{\cal C \in N^-}\;M_{\cal C} P_{\cal C} \right) + \left( \sum_{\cal C \in P^+}\;M_{\cal C} P_{\cal C} - \sum_{\cal C \in P^-}\;M_{\cal C} P_{\cal C} \right) \\
= \left( \sum_{\cal C \in N}\;M_{\cal C} P_{\cal C} - \sum_{\cal C \in P}\;M_{\cal C} P_{\cal C} \right) + 2 \left( \sum_{\cal C \in P^+}\;M_{\cal C} P_{\cal C} - \sum_{\cal C \in N^-}\;M_{\cal C} P_{\cal C} \right)\\
\ge \left( \sum_{\cal C \in N}\;M_{\cal C} P_{\cal C} - \sum_{\cal C \in P}\;M_{\cal C} P_{\cal C} \right) = 1 - \frac{2}{N} \left( \sum_{\cal C \in N}\;(N_-)_{\cal C} P_{\cal C} + \sum_{\cal C \in P}\;(N_+)_{\cal C} P_{\cal C} \right)\\
\ge 1-\frac{4}{N} \sum_{p\in V} \sum_{b_{min}} \mu^p(b_{min})\;e^{-2 \beta J_{min} b_{min}}\\
\ge 1- 4 e^{-2\beta J_{min}} \sum_{b\ge 1} \mu(b_{min})\,e^{-2\beta (b-1) J_{min}},
\end{multline*}
When the sum on the last line converges, the equation tells us that, for large enough $\beta$ (low temperatures), $\langle |M|\rangle$ is greater than a positive constant, so that spontaneous magnetization on an infinite graph is achieved, while in general $\langle |M| \rangle$ is finite for every $N$, but can tend to zero as $N\rightarrow\infty$.
\end{proof}
The main problem in employing the previous theorem is determining bounds on $\mu(b)$.
The simplest case in which the hypothesis does not hold is a situation in which for some finite $b$ the number of borders surrounding a given point is infinite. As a sound check of the validity of the theorem, all the \emph{weakly separable}\cite{camparicassi2009} graphs, which do not magnetize, belong to this category.\par
To further our understanding of the result, we need to present a new parameter.
Given a subgraph $A\subset {\cal G}$, we define its external boundary $\partial A$ as the set of points in ${\cal G}\setminus A$ that have a bond to a point in $A$; denoting the number of vertices in $A$ as $\left\vert A \right\vert$ we now present the {\bf isoperimetric dimension} $d_{iso}$ as the minimum $d$ such that $\partial A \ge C\cdotp \left\vert A \right\vert^{d -1\over d}$. The largest set of points encompassable with $b$ links is thus smaller than $b^{d_{iso}\over d_{iso}-1}$; since these points are connected, $b^{d_{iso}\over d_{iso}-1}$ is also the maximum radius of a set including $i$ with a border $b$, so the set of reachable points, $V(b)\subset {\cal P}$, has a cardinality $\left\vert V(b)\right\vert \le b^{d_{iso} d_{frac}\over d_{iso}-1}$.\par
Given a point $p$ and for each border $\cal B$, consider $\bar{\cal B}$, the collection of vertices contributing to $\cal B$ which are on the inside of $\cal B$ with respect to $p$. As the two are in biunivocal relation once $p$ is chosen, counting the borders is the same as counting the vertex borders.

As a consequence of the previous paragraphs, the following holds:
\begin{proposition}
In a graph with isoperimetric dimension $d_{iso}>1$, the number of possible borders surrounding $p$ is bounded by
\[ \mu^p(b) \le \sum_q^{\left\vert V(b)\right\vert} N_q(b) \le b^{d_{iso} d_{frac}\over d_{iso}-1} \cdotp N_{Sup}(b), \]
where the sum is over the points $q$ which can be enclosed in a border of size $b$, $N_q(b)$ is the maximum number of vertex borders of length $b$, containing $p$, which can be created starting from the point $q$, and $N_{Sup}(b)=\sup_{q\in V(b)} N^p(b)$.
\end{proposition}

A border is \emph{connected} if the corresponding vertex border is a connected set.
We will need the following proposition regarding $N^p(b)$ to obtain a general result.
\begin{proposition}
The number of connected vertex borders starting from a given point $p$ grows at most exponentially with $b$: $N^p(b)\le const\cdotp C^b$.
\end{proposition}
\begin{proof}
A {\bf tree} is a graph that has no loops.
From each connected subgraph $A$ we can draw a number of different spanning trees, i.e. trees having the same set of points $\cal P$ as the original $A$. For any spanning tree we can construct a path visiting all its vertices in no less than $b-1$ steps, and in no more than $2 z_{Max} b$ steps.
While the former statement is obvious, we now prove the latter using the following algorithm:
starting from $i$, choose link and cross it; at each vertex on the path, choose a link not yet crossed; if there is no free link, step back through the link from which the path first arrived at the vertex. With this algorithm, each link is crossed no more than twice, so the path is of no more than $2 z_{Max} b$; furthermore, all the links are crossed, so, since the boundary is connected, all the points are visited.
As a consequence, all the spanning trees of $b$ points starting from a vertex $i$ can be constructed as paths of $b-1,b,\dots,2 z_{Max} b$ steps. Since each step can be chosen among at most $z_{Max}$ links, the total number of possible spanning trees, starting from $i$ and made of $b$ vertices, is less than
\[ z_{Max}^{b-1} + z_{Max}^{b} + \dots + z_{Max}^{2 z_{Max} b} = { z_{Max}^{2 z_{Max} b + 1 } - z_{Max}^{b-1} \over z_{Max}-1} , \]
and the thesis follows:
\[ N^p(b) \le const \cdotp C^b . \]
\end{proof}

When a vertex border is made of more disconnected parts, $\mu(b)$ grows exponentially if, for all borders, it's possible to connect all the parts using no more than $l\cdotp b$ vertices, where $l$ is a constant of the graph: in fact in this case to each border of length $b$ corresponds one connected vertex border of length between $b$ and $b\cdotp l$, so that
\[ N(b) \le C^{b}+C^{b+1}+\dots+C^{l\cdotp b} = {C^{ l\cdotp b + 1 }  - C^b\over C-1}.\]
\begin{figure}
 \includegraphics{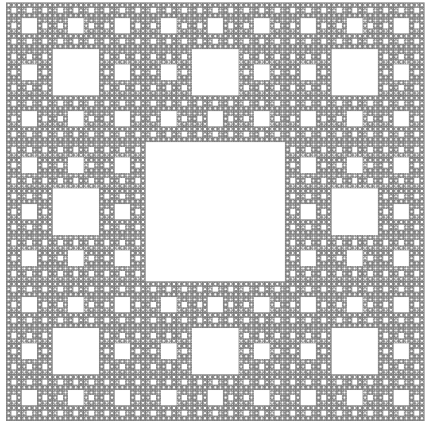}%
 \caption{\label{sierpinskicarpet} The Sierpinski carpet allows for disconnected borders, but they can be connected with no more with $l\cdotp b$ links, so the number of possible borders grows no faster than an exponential with $b$, and the graph magnetizes.}
\end{figure}

Noting that $d_{iso}>1$ implies that there is no border length $b$ for which $\mu^p(b)$ is infinite, the previous results can be combined to form the following theorem.
\begin{theorem}
For all graphs with isoperimetric dimension $d_{iso} > 1$ and vertex borders which are connectable with no more than $l\cdotp b$ vertices, a finite critical $\beta_c<\infty$ exists such that for all $\beta>\beta_c$ spontaneous magnetization is achieved.
\end{theorem}
To the latter category belong the regular lattices in $d\ge 2$ dimensions and crystals with any kind of elementary cells; we explicitly note that for an Euclidean lattice in $d=2$ dimensions we recover the result by Griffiths\cite{griffithspeierls}.
In addition, each vertex border can be connected with no more than $l\cdotp b$ vertices in the Sierpinski carpet too, which therefore magnetizes, in accord with the existing literature\cite{Vezzani2003} (see Fig.\ref{sierpinskicarpet}).
\par
\begin{figure}
 \includegraphics{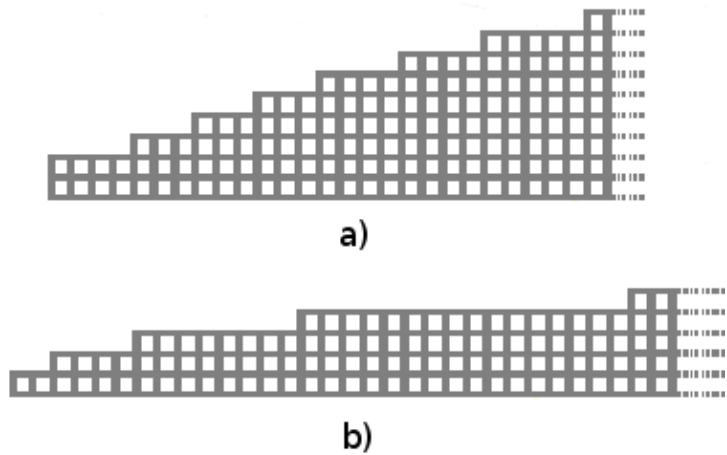}%
 \caption{\label{growingladder} Two examples of growing ladder graphs: in (a) all the borders are connected (remember that, following the definition we use here, a border divides a graph into exactly two subgraphs, each \emph{connected}), and as a consequence the graph will magnetize. In (b) too the borders are connected, and even if the growth in width of the ladder is very slow (logarithmic) the same result holds.}
\end{figure}

Consider now the ladders of infinitely growing height (see Fig.\ref{growingladder}): they are structures described, at any offset $n$ on the semi-infinite base line, by a non-decreasing integer function $h(n)$; as long as the isoperimetric dimension of the ladder is strictly greater than one ($h(n)\ge A_0 n^\alpha + B_0$, for $\alpha>0$, $A_0>0$ and $B_0>0$) the previous arguments apply, so the ladder magnetizes; on the other hand, when $d_{iso}=1$ a little more work is required: the total number $V(b)$ of vertices which can be included in at least one border of length $b$ can not grow faster than the number of points on the left of the rightmost border of length $b$; the latter is at at offset $n(b)=\max\{ n' \vert b=h^{-1}(n') \}$, so that $V(b)$ satisfies $V(b) \sim \sum_{i=1}^{n(b)} h(i)$ for large $b$;
when
\[ B_0 + A_0 \log i \le h(i) \le B_1 + A_1 i^{\alpha} \]
for some $A_0,A_1 >0$, $B_0,B_1\ge 0$ and $\alpha\ge1$, the volume satisfies
\[ V(b) \le \int_1^{e^{b/A_0}} di\;\left( B_1 + A_1 i^\alpha \right) \sim {b\over A_0} {A_1\over 2 \left( \alpha+1 \right)} \; e^{2\left( \alpha+1 \right)b/A_0},\]
and since the borders are all connected $\mu(b)$ is exponential and the graph magnetizes.
When instead $\lim_{n\rightarrow\infty} {h(n)\over \log n} = 0$, for all $\epsilon>0$ and $n$ large enough $h(n)\le\epsilon \log n$ holds; as a consequence,
\[ V(b) \ge \int_0^{b\over\epsilon} di\;h(i) \;> \int_0^{b\over\epsilon}di = e^{b\over\epsilon} \]
holds for all $\epsilon>0$; in this case the sum $\sum_b \mu(b) e^{-2\beta \left( b-1 \right) J_{min}}$ diverges for all temperatures, so the hypotheses of our theorem are not fulfilled.
These results are in agreement with a result by Chayes and Chayes\cite{Chayes1986} about more general structures called $d$-wedges, where it is proved that $h(n)\ge \log n$ is both a sufficient and a necessary condition for spontaneous magnetization.
\par

\section*{Discussion}
To give a more intuitive interpretation of the theorem we proved, we can proceed as follows: if the number of borders grows less than exponentially, we can argue that all of these borders will contain a number of spins increasing slowly with the length of the border; as a consequence, the formation of large
clusters of spins in a magnetized graph will be energetically unfavoured, so that the latter will result a stable state.
On the other hand, if $\mu^i(b)$ grows very fast with $b$, we expect that some of the borders will be far from the vertex $i$, so that more and more vertices will be enclosed in short (low $b$) borders; this in turn means that large clusters of spins can be flipped spending a small amount of energy, so that a magnetized graph may be unstable with regard to thermal fluctuations.\par
The condition of our theorem is a strong one, in that it investigates a global property of the graph. For this reason it can not be a necessary condition for achieving spontaneous magnetization: if a graph has a part, which has zero measure in the thermodynamic limit, for which the number of boundaries $\mu^p(b)$ is greater than any exponential (e.g. a semi-infinite line connected to a point on a plane), the hypothesis of the theorem is false but the graph as a whole can still magnetize.\par
An important, yet straightforward, observation is that whenever a subgraph of non zero measure exists that is magnetizable, all the graph is magnetizable: in fact all the correlation functions, as computed on the subgraph, are smaller than or equal to the corresponding ones in the complete graph; when, on the other hand, the graph is formed by a collection of zero measure, weakly connected, magnetizable subgraphs (e.g. an infinite collection of parallel planes, each connected via a single link to the next one), there is no guarantee that $\langle \vert M \vert \rangle>0$.\par
Our result about the Ising model on graphs is a further step towards a full comprehension of the mechanism of phase transitions on general networks: together with a sufficient condition for the lack of spontaneous magnetization\cite{camparicassi2009}, it allows to ascertain the magnetizability of a large number of structures with a minimal amount of computation.\par
Further steps extending this work should aim at closing the gap between magnetizable and non-magnetizable graphs under the $\langle \vert M \vert\rangle$ definition, in order to identify a condition both necessary and sufficient for spontaneous magnetization; another direction of development could be to treat non symmetric situations, as in Pirogov-Sinai theory.
\bibliographystyle{apsrev4-1}

\end{document}